# Developing Intellectual Network Management Facilities by Means of Pattern Recognition Theory

*Yuriy A. Chashkov*

Thesis main purpose consist in LAN management intellectualization methods and facilities developing. To achieve this aim following problems was solved:

- Existing network management software was analyzed;
- Network management architecture and models was analyzed;
- Requirements to intellectual network management system was developed;
- Gathering and preliminary information processing methods was developed;
- Network managerial process intellectualization models, methods and software was developed.

## I. WORK DESCRIPTION

**The first chapter** of dissertational work is devoted to the detailed analysis of engineering environment, existing models and network management systems architecture. Contains management problems classification from the point of view necessary network parameters maintenance. In this chapter the main problem arising in network equipment management - insufficiency aprioristic information on its functioning also is determined. In result it is shown, that management systems based on traditional necessitation approach, are ineffective in dynamic multiparametric poorly determined network equipment management. Therefore development of new approaches in realization of networks management automation problem is required. It is shown, that with enough plenty of network management systems, exist lack of Intellectual Control Facilities which allow us not only collect network statistics and channels loading intensity observing, but also analyze gathered information by means of Artificial Intellect that gives an opportunity more effectively utilize network resources. In LAN management it is necessary to solve next tasks:

- Control period definitions under functioning characteristics;
- Gathering and transfer of the information about controlling objects;
- Information processing and acceptance managing decisions;
- Managing influence realization.

At such approach LAN management is informational-analytical process. It consider some set of gathering, transferring, storage and analyzing procedures of LAN state information. The primary managing structure goal consists in LAN equipment state identification on the received data on his state and comparison it with preference mode, also preparation and realization managing influence.

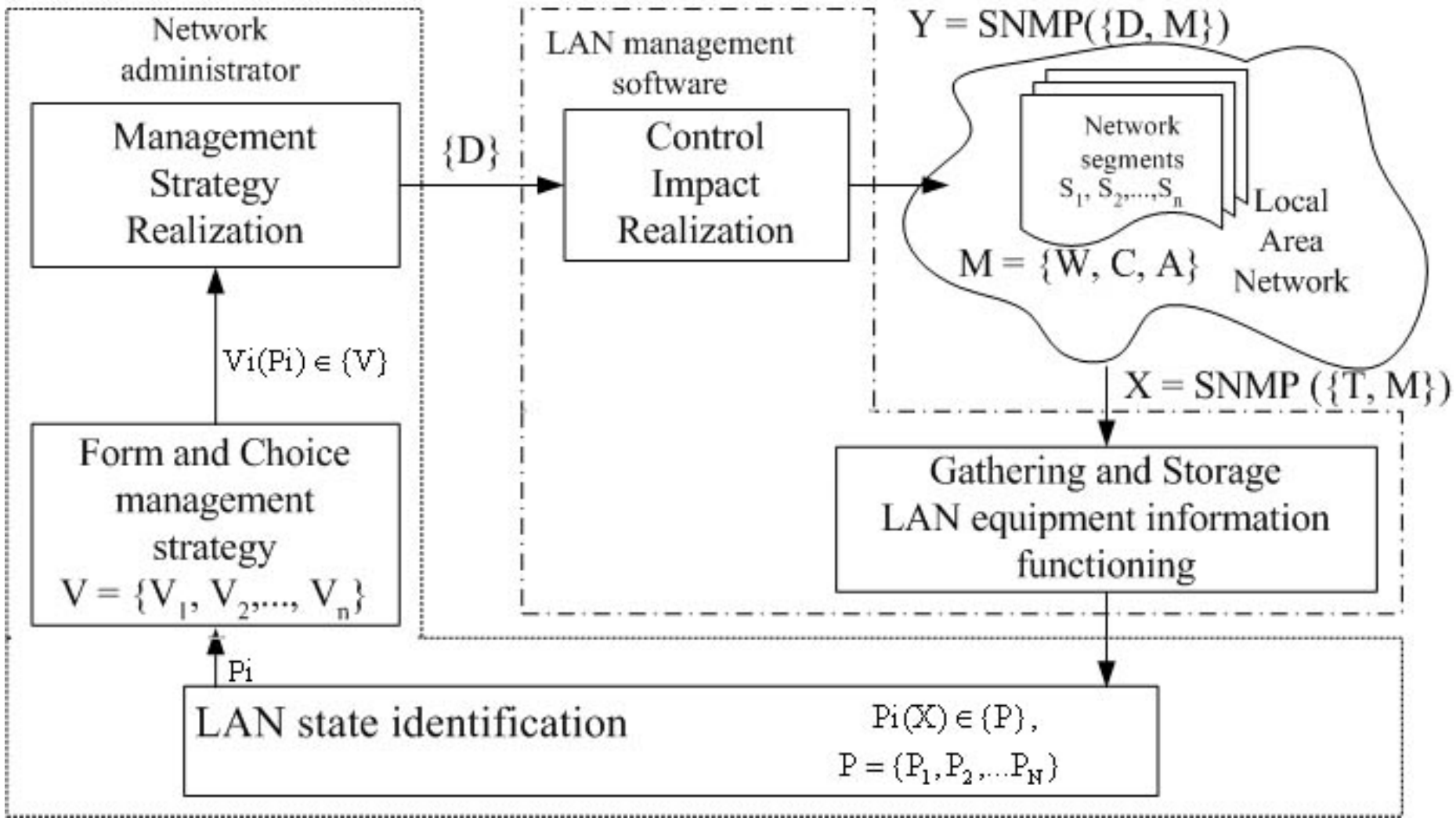

Fig.1. Intellectual Network Management System Model

S1, S2,...,Sn - Network segments;
M - Management objects;
W - Workstations;
C - LAN servers;
A - Active telecommunication equipment;
X - Parameters which describe current network state;
T - Time;
P - Possible state;
V - LAN management strategies;
D - MIB parameters , which define management object functionining;
Y - Control impact realizaion function.

On fig.1. LAN intellectual management system model is given. On the basis of LAN state received information, and available statistics LAN administrator makes decision about LAN equipment state therefore this or that management algorithm out is formed.

Pattern Recognition is not that other, as object state identification. On fig. 2, comparative analysis of Pattern Recognition and LAN managements procedures is given. It is visible, that these tasks are identical. Hence, the opportunity of pattern recognition system application at LAN equipment state identification stage is represented quite obvious.

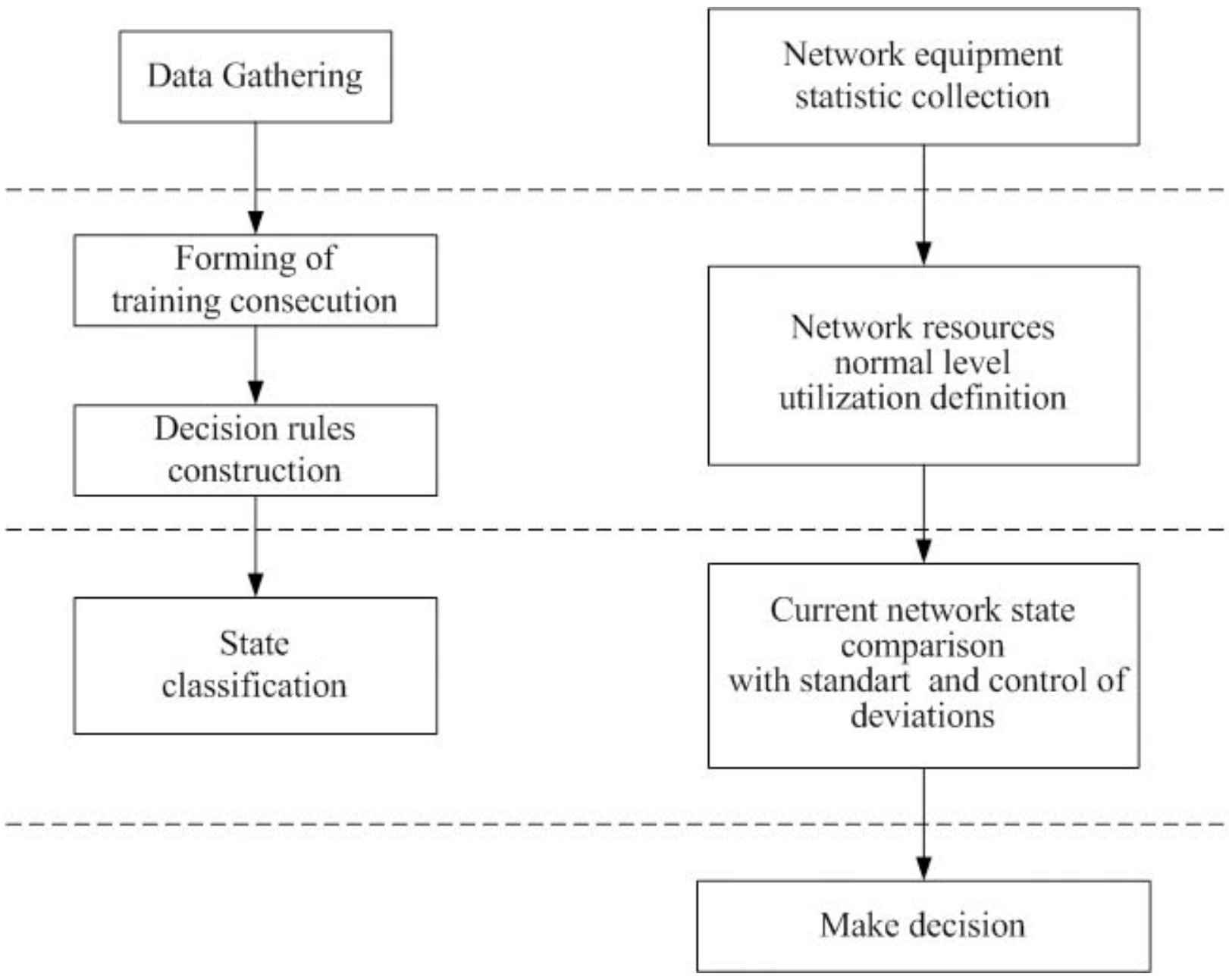

Fig. 2. Pattern Recognition and LAN management procedures comparative analysis

**In the second chapter** the complex approach to network equipment status identification procedure realization as dynamic object with use Pattern Recognition theory is considered. Network managerial process formal description in terms of Pattern Recognition theory is given. As a result of the carried out analysis of existing Pattern Recognition methods, and their using opportunities and realization in Intellectual Network Management System (INMS) solved to use Potential Function method, in deterministical statement which realize recurrent procedure of average risk minimization idea.

Potential Functions method assumes existence in space $X$ system $\varphi_i(x)$, $(i = 1, 2,..., N)$, allowing for everyone pair of divided sets to find such number at which anyone class $Q$ in training sequence characterized by total potential function:

$$K(q, y) = \sum_{x_j \in Q} c_{qj} \varphi(x_j, y),$$

where summation conducted on all objects belonging to class $Q$. The number $c_{qj}$ represents weight factor object $j$ belonging to class $Q$ of training sequence and arise during training. It is considered, that any vector belongs to class $Q$, if $K(q, x_j) = \max_p K(p, x_j)$.

Training consists of two stages. At **the first stage** there is a set of deciding vectors $S_p$, and constants $c_p$ such, that $[(S_q, x) + c_q c_0] - [(S_p, x) + c_p c_0] > \varepsilon$, where $p=1...K;$ $q=1...K;$ $p \neq q;$ for all $x$, belonging to class $Q$. Here $\varepsilon$ defines division border: if $\varepsilon\neq 0$ classes division border set in interval $[+\varepsilon, -\varepsilon]$ and vectors which have got in this interval, will be considered unidentified (**it is possible to specify belonging to the certain class**).

Determining condition, that on step $r$ is necessary to make weight factors reorganization, it is set by an inequality:

$$\left(\sum_{x_q\in P} a_{pq}^{r-1} f(x_q, x_r) - \sum_{x_q\in G} a_{gq}^{r-1} f(x_q, x_r)\right) \le \varepsilon$$

where $\varepsilon\geq 0;\ G\neq P;\ x_r$ – object $j$ from training sequence, presented to the machine at step $r$. Vector $x_r$ belongs to class $P$; $c_{pq}^{r-1}$ - vector $q$ weight factor value from class $P$, calculated at previous step.

If this condition is not carried out, weight factors remain without changes and next vector of training sequence showing to the machine. There are two variants of weight reorganization factors on step $r$:

a) $c_{pj}^{r} = c_{pj}^{r-1} + \Delta; \quad \Delta > 0,$

$c_{gj}^{r} = c_{gj}^{r-1} \quad for \quad g \neq p,$ g = 1...K;

b) $c_{pj}^{r} = c_{pj}^{r-1} + \Delta, \quad c_{gj}^{r} = c_{gj}^{r-1} - \Delta,$

where $c_{gj}^{r-1}$ determining from a ratio:

$$\sum_{x_q\in G} c_{gq} f(x_q, x_j) = \max_p \left\{ \sum_{x_q\in p} c_{pq}^{r-1} f(x_q, x_j) \right\}$$

Training comes to end, when at the next training sequence reviewing all vectors identified correctly or after preset viewings number.

At **the second stage** it is supposed, that machine except training sequence contains $c_p$ constants (number of vectors concerning to class $P$) and decision functions $S_p$ received at the first stage are entered. At this stage training sequence is looked through some times; thus each time values $S_p$ and compare are corrected. Condition of their reorganization on step $r$ determined by expression:

$$\Delta_p = \max_q \left\{ \frac{(c_q^{r-1}-1)K_q(x_j) - 2c_q^{r-1}S_q^{r-1}}{(c_q^{r-1}+1)c_q^{r-1}(c_q^{r+1}-1)} - \frac{(c_i^{r-1}-1)K_i(x_j) - 2(c_i^{r-1}-1)S_i^{r-1}}{c_i^{r-1}(c_i^{r-1}-1)(c_i^{r+1}-2)} \right\}$$

where $q=1...K$; $c_q^{r-1}$ - number of vectors referred to class $Q$ at (r-1) step, $S_q^{r-1}$ - class $Q$ potential function at $(r\text{-}1)$ step, $i$ - class index to which vector $X_j$ was referred at $(r\text{-}1)$ step. If this condition is carried out, under formulas:

$$S_p^r = S_p^{r-1} + K_p(x_j), \quad c_p^r = c_p^{r-1} + 1;$$
$$S_i^r = S_i^{r-1} - K_i(x_j), \quad c_i^r = c_i^{r-1} - 1$$

decision functions reorganization is carried out.

If all vectors from training sequence at the next reviewing are distinguished precisely the same as at previous time or training comes to end. Training also comes to end in case of training sequence was seen preset times number.

At work in reorganization mode in memory should be:

- recognized vectors sequence $x_1, x_2, ...x_L, x_{L+1}, ..., x_{L+g}$;
- potential functions $S_p$, constructed in training mode;
- sequence of constants $c_p$ vectors determining quantity referred to each of classes;
- classes indexes of each training sequence vectors $x_1, x_2, ...x_L$, received in training mode;

In practice of algorithm using there is a problem of Potential Function $U(x, x^*)$ choice. In some cases it is not necessary to care of a functions $\varphi_i(x)$ and constants $\alpha_i$ preliminary choice. At once it is possible to set a kind of potential such that function would be enough smooth, and accepted maximal value at $x=x^*$. Taking into account that more difficultly class of deciding rules, is more probability to receive bad rule at recognition of new conditions. Therefore, taking into account all aforesaid, and being based on spent before theoretical and practical researches, we in the given work as function for determining "potential" between $q$ and $j$ conditions shall use:

$$f(x_q, x_j) = \frac{1}{1 + \alpha R_{qj}^2}$$

where $\alpha > 0$; $R_{qj}^2 = \sum_{i=1}^{n} (x_j^2 - x_q^2)^2$ - evclid distance between $q$ and $j$ conditions; $n$ – vector $x$ dimension.

**In the third chapter** INMS generalized model is offered and proved. Construction principles and the basic INMS functional components are determined. The basic requirements which are formulated should correspond to INMS. Network management subsystem generalized circuit which used in INMS shown in fig. 3.

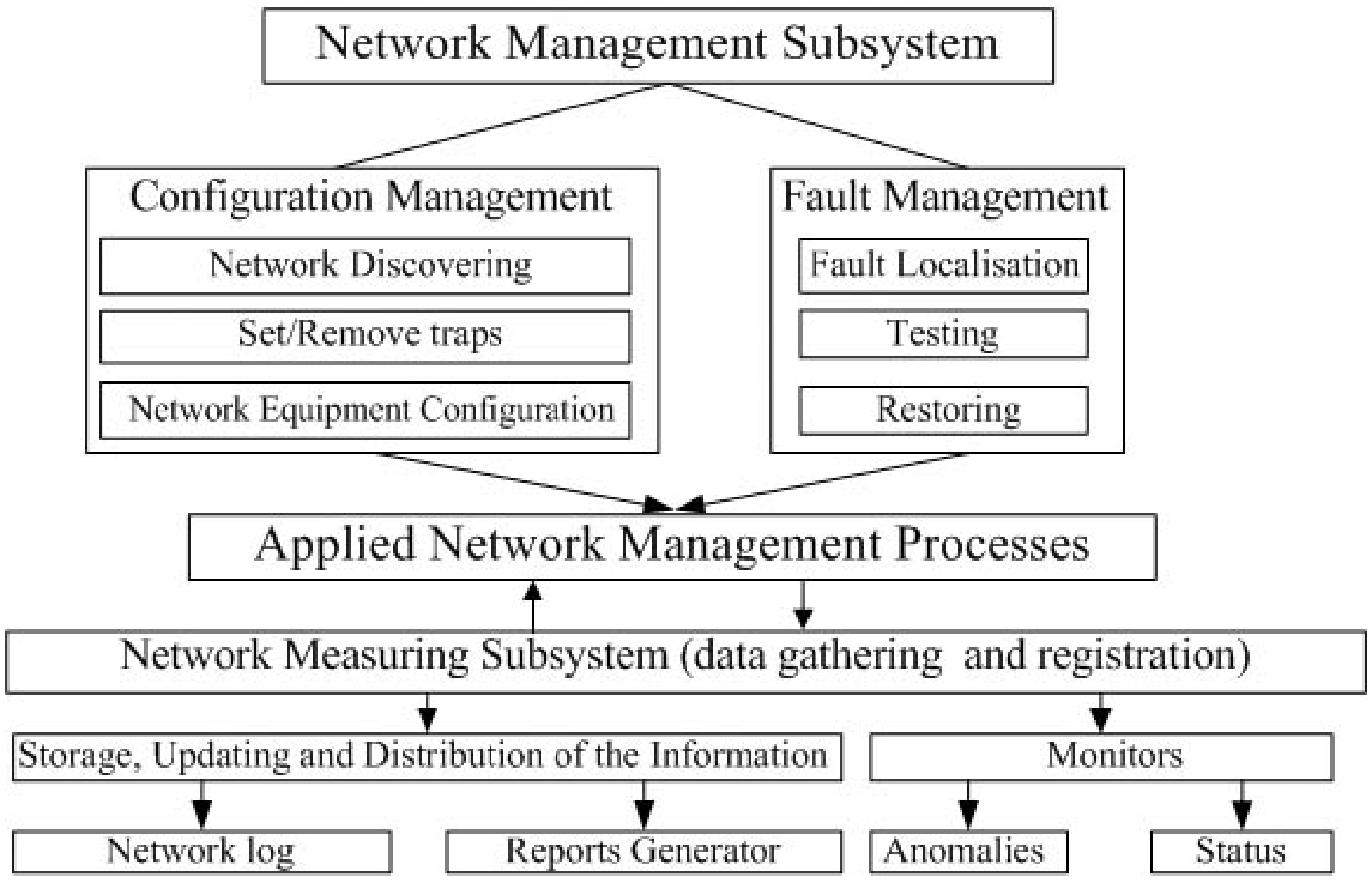

Fig. 3. Network management subsystem generalized block diagram

Apparently from the given circuit, all functional components lean on the common network measuring subsystem (data registration and information gathering). Doubling same functions in other subsystems in this case excluded, and the common data can be used for various purposes.

**In the fourth chapter** problems of practical developing INMS "AINetMon" are solved on the basis of the theoretical positions obtained in the previous chapters. Developed software «AINetMon» represents automated structure which realize operative intellectual network management. This management is carried out due to gathering, processing, and network equipment statistic storage from appropriate MIB's, and also due to regular procedure of current status classification by means of pattern recognition method which realize network management intellectualization procedure.

This program is written as multithread application by means of Java Developer Kit 1.3 and specialized libraries - AdventNet. Program realized by means of RMI technology as several cooperating streams. It has allowed to create the distributed network management system most effectively use system resources and has given opportunity of real time information processing. At practical realization developed theoretical methods was decide to use Web-technology together with JAVA language as a basis for construction management system. Such technical solution has allowed us to create platform-independent distributed management system, potentially capable overcome SNMP protocol lack.

## II. SCIENTIFIC AND PRACTICAL RESULTS

The primary Thesis **scientific results** are:

- INMS generalized model;
- Pattern Recognition theory using in active network equipment state identification;
- Intellectual network management software generalized structure which realize offered methods;
- Intellectual network management software "AINetMon" that realized developed models and methods.

Thesis **practical importance** consists in:

1. Opportunities of use dissertational work results by development, adjustment to concrete conditions, and in the solution of active network equipment productivity analyzing problem.
2. Practice using developed methods, intellectual network management architecture, and conceptual designing technique allowed rationally design and realize applied data processing algorithms as a high-grade INMS which appreciably allows facilitate networks administration and more effectively predict possible malfunctions.
3. Using of object-oriented language JAVA, modern network technologies based on SNMP protocol, MIB structures and AdventNetSnmpv3 libraries as a basis for construction INMS has given to created system platform independence.
4. Creation software for network management intellectualization common problem solution, including about 90 classes that covers set of problems: gathering and initial statistical data processing, intellectual component training, network status classification and history storage, efficiency maintenance of managing solutions acceptance. Displaying received results in the convenient and evident form.

Thesis results implantation. Practical realization of developed software has allowed to facilitate LAN management, effectively predict possible malfunctions, and reduce time for their search and elimination. Thesis main practical results are implanted in the next organizations: Moscow State Institute of Electronics and Mathematics, Chernovtsy National University, State Research-and-Production Enterprise «Chernovtsy GeoInfoCentr», LantaTechService ltd, Joint-Stock Company «ALP-96».